\documentclass[a4paper,11pt]{article}
 
 \usepackage[margin=1.5in]{geometry}    
 \usepackage[english]{babel}
 \usepackage{tikz}
 \usepackage{here}
 \usetikzlibrary{positioning}
 \usepackage{amsmath, amssymb}
 \usepackage{booktabs}
 \usepackage{hyperref}
 \usepackage{fancyvrb}
 \usepackage{fvextra}
 \usepackage{multicol}
 \title{
     Symbolic, numeric and quantum computation of Hartree-Fock equation
 }
 
 \author{Ichio Kikuchi$^{1}$, Akihito Kikuchi$^{2}$\footnote{akihito\_kikuchi@gakushikai.jp (The corresponding author; a visiting researcher in IFQT)} 
\\
         \small $^{1}$Internationales Forschungszentrum f\"ur Quantentechnik\\
         \small $^{2}$International Research Center for Quantum Technology, Tokyo \\
 }
 
 \date{\today}    

\begin{document}\maketitle
 \begin{abstract}
 
 In this article, we discuss how a kind of hybrid computation, which employs symbolic, numeric, classic, and quantum algorithms,  allows us to conduct Hartree-Fock electronic structure computation of molecules. In the proposed algorithm, we replace the Hartree-Fock equations with a set of equations composed of multivariate polynomials. We transform those polynomials to the corresponding Gr\"obner bases, and then we investigate the corresponding quotient ring, wherein the orbital energies, the LCAO coefficients, or the atomic coordinates are represented by the variables in the ring. In this quotient ring, the variables generate the transformation matrices that represent the multiplication with the monomial bases, and the eigenvalues of those matrices compose the roots of the equation. The quantum phase estimation (QPE) algorithm enables us to record those roots in the quantum states, which would be used in the input data for more advanced and more accurate quantum computations.
 
 \end{abstract}
 
 \section{
 Introduction
 }

 There is a symbolic-numeric method of quantum chemistry \cite{KIKUCHI2013}, whereby the computations are carried out in the following way:
 
 \begin{itemize}

 \item 

The molecular integrals are represented by the polynomial approximation of analytic formulas, which are computed symbolically if we use analytic atomic bases, such as Gaussian-Type or Slater-Type orbitals (GTO or STO) \cite{szabo2012modern}. Those formulas are the analytic functions of several variables, namely, those of orbital exponents and atomic coordinates. By Tailor expansion with respect to those variables, the molecular integrals are approximated by polynomials.
 
 \item The total energy is a polynomial composed of molecular integrals and the undetermined coefficients of LCAO. The ortho-normalization conditions are similarly treated.
 
 \item We compose the objective function from the total energy and the ortho-normalization condition with the Lagrange multipliers which represent the orbital energies.
 
 \item By symbolic differentiation, we obtain a system of polynomial equations that gives the optima.
 
 \item To get the roots of the system of polynomial equations, we apply several methods of computer algebra, where Gr\"obner bases and the primary ideal decomposition play central roles in getting the quantum eigenstates \cite{decker2006computing, SOTTILE,cox2006using,cox2013ideals, ene2011gr, eisenbud2013commutative}. Namely, we compose an ideal $I$ from the given polynomials and transform them into another system that has a more suitable form for root-finding \cite{gianni1988grobner, eisenbud1992direct,moller1993decomposing,primdecSY,lazard1992solving,moller1993decomposing}. The ideal representing a Hartree-Fock equation could be decomposed into several subsystems described by primary ideals. Each primary ideal would represent one solution set, namely, one quantum state if the decomposition is executed to the full.
 
 \item Up to now, we have reported the results of several simple molecules, using STO and n-GTO models ( \cite{KIKUCHI2013} and \cite{kikuchi2022ab}).  In those works, the adopted algorithms are classical, not quantum.  One might use the term \textit{Molecular Algebraic Geometry} to refer to this algebraic computational scheme for molecular orbital theory.
 
 
 \end{itemize}

 The algebraic method described above could relate to the quantum algorithms, and the theme of the present study is to demonstrate it. This article is structured as follows.  First, we show the computational step whereby the classical symbolic computation prepares the eigenvalue problem that gives the roots of the given equation. Second, we show how the quantum algorithm could solve the problem. Then we discuss several points that should be treated with care.
 
 \section{Computational process}
 In this section, we incarnate the algorithms for symbolic-numeric and classical-quantum computation in quantum chemistry.  The computational process is composed of two phases. The first phase uses the symbolic-numeric classical algorithm and converts the Hartree-Fock equations into a representation suitable for quantum computation. The second phase uses the data generated in the first phase and computes the roots of the Hartree-Fock equations. We describe the algorithm in each phase, using two examples.
 
 \subsection{Phase 1: symbolic-numeric classical algorithm}
 \subsubsection{Tools for symbolic, numeric, and classical computation}
 
 We solve the set of polynomial equations through the computational steps explained in \cite{sottile2002enumerative}. 
 
 \begin{itemize}
 
 \item Let $I$ be an ideal made of multivariate polynomials $(f_1, f_2,...,f_{t})$ in $R[x_1,x_2,...,x_n]$. Once the Gr\"obner for the ideal $I$  is computed, it is an easy task to represent any element in $R[x_1,x_2,...,x_n]/I$ uniquely as the linear combination of the monomial basis of the quotient ring.
 
 \item Let $\bar{x}_1,\bar{x}_2,...,\bar{x}_2$ 
be the representatives of $x_1,x_2,...,x_n$ in $R[x_1,x_2,...,x_n]/ I$. Additionally, let $b$ be a vector that is composed of the representatives of the monomial basis of the quotient ring.
 
 \item For any $i$, the multiplication $\bar{x}_i \cdot b$ is represented by
 \begin{align}
 \bar{x}_i \cdot b  = b \cdot M_{x_i}
 \end{align}
 with a transformation matrix $M_{x_i}$. The entries of the matrix are numbers, but not symbols.
 
 \item As $M_{x_i}\cdot M_{x_j}=M_{x_j}\cdot M_{x_i}$, those transformation matrices share common eigenvectors $\{v_j |j=1,...,M\}$, where $M$ is size of the monomial basis $b$.
 
 \item Let us consider the eigenvalue problems, defined as follows,
 \begin{align}
 \bar{\xi}_i^{(j)} \ v_j  = v_j \cdot M_{x_i}
 \end{align}
 for $i=1,...,n$ and $j=1,...,M$. Those equations are solved numerically, and the eigenvalues give the common zeros of the polynomials included in the ideal $I$. Namely, the eigenvalues give the roots of the set of polynomial equations defined by
 \begin{align}
 f_1(x_1,x_2,...,x_n)=f_2(x_1,x_2,...,x_n)=...=f_t(x_1,x_2,...,x_n)=0
 \end{align}
 in such a way that
 \begin{align}
 (x_1,x_2,...,x_n)=(\bar{\xi}_1^{(j)},\bar{\xi}_2^{(j)},...,\bar{\xi}_n^{(j)})
 \end{align}
 for $j=1,...,M$. Note that if eigenvectors $\{v_j\}_j$ for one $M_i$ is obtained, the other components of the roots are computed by
 \begin{align}
 \bar{\xi}_i^{(j)}= \frac{(v_j \cdot M_{x_i},v_j)}{ (v_j ,v_j)}.
 \end{align}
 \end{itemize}
 
 The root-finding of a system of polynomial equations is replaced by a set of eigenvalue problems, which could be solved by quantum algorithms. We put the eigenvectors $\{v_j\}_j$ into a set of quantum states $\{| v_j\rangle\}$, and the computational steps are carried out by a quantum circuit, which conducts the following transformation:
 \begin{align}
 | v_j\rangle |Ancilla_1\rangle
 |Ancilla_2\rangle \cdots
 |Ancilla_n\rangle
 \rightarrow | v_j\rangle 
 |\bar{\xi}_1^{(j)}\rangle |\bar{\xi}_2^{(j)}\rangle \cdots |\bar{\xi}_n^{(j)}\rangle
 \end{align}
 where the eigenvalues of $M_{x_i}$ $(i=1,...,n)$ for $v_j$ are recorded in ancilla qubits through a successive application of quantum phase estimation.

 \subsubsection{Computation for a simple toy model}
 Let us compute a simple toy model, where the secular equation is given by
 \begin{align}
 \begin{pmatrix}
     V(x,y) & -1\\
     -1 
& V(x,y)
 \end{pmatrix} 
 \begin{pmatrix}
     x \\
     y 
 \end{pmatrix} 
 =e\begin{pmatrix}
     x \\
     y 
 \end{pmatrix}
 \end{align}
 along with the normalization condition
 $x^2+y^2=1$. 
 The variables $(x,y)$ are the amplitudes of the wavefunction and $e$ is the orbital energy. $V(x, y)$ is the on-site potential that is the function of the amplitude of the wavefunction. We assume that the roots are real. 
 
 The polynomial ideal that represents the secular equation is given by
 \begin{align}
 I=(x V(x,y)-y - e\,x, y V(x,y)-x - e\,y,x^2+y^2-1)
 \end{align}
 
 In the case of $V(x,y)=0$, the Gr\"obner basis is given by
 \begin{align}
 I_{std}=(e^2-1,2y^2-1,x+ye)
 \end{align}
 The roots of the set of polynomial equations are given by
 \begin{align}
 (x,y,e)=\left(\pm\frac{1}{\sqrt{2}},\pm\frac{1}{\sqrt{2}},-1\right),\left(\pm\frac{1}{\sqrt{2}},\mp\frac{1}{\sqrt{2}},1\right)
 \end{align}
 
 The entries in the quotient ring $Q(x,y,e)/ I$ are the linear combinations  of the monomial basis $b=(b[0],b[1],b[2],b[3])$:
 \begin{align}
 b[0]&=y\,e \\
 b[1]&=y \\
 b[2]&=e \\
 b[3]&=1
 \end{align}
 
 In the quotient ring, the multiplications of the entries of the basis $b$ by $x$, $y$, and $e$ are represented by the transformation matrices: $b\cdot p=b \cdot m_p $ for $p=x,y,e$.
 \begin{align}
 m_x=
 \begin{pmatrix}
     0 
& 0 & 0 &-1\\
     0 
& 0 & -1 & 0\\
     0 & -0.5 & 0 & 0 \\
     -0.5 & 0 & 0 & 0 \\
 \end{pmatrix} 
 \end{align}
 
 \begin{align}
 m_y=
 \begin{pmatrix}
     0 
& 0 & 1 & 0\\
     0 
& 0 & 0 & 1\\
     0.5 & 0 & 0 & 0 \\
     0 & 0.5 & 0 & 0 \\
 \end{pmatrix} 
 \end{align}
 
 \begin{align}
 m_e=
 \begin{pmatrix}
     0 
& 1 &  & 0\\
     1 
& 0 & 0 & 0\\
     0 & 0 & 0 & 1 \\
     0 & 0 & 1 & 0 \\
 \end{pmatrix} 
 \end{align}

 
 For the above example, the related properties are given as follows.
 \begin{align}
 \begin{pmatrix}
 v & (v m_x, v) & (v m_y ,v) & (v m_e, v) \\ 
 (-\frac{1}{\sqrt{2}},\frac{1}{\sqrt{2}},-1,1) & \frac{1}{\sqrt{2}} & \frac{1}{\sqrt{2}} & -1 \\ 
 (\frac{1}{\sqrt{2}},-\frac{1}{\sqrt{2}},-1,1) & -\frac{1}{\sqrt{2}} & -\frac{1}{\sqrt{2}} & -1 \\ 
 (\frac{1}{\sqrt{2}},\frac{1}{\sqrt{2}},1,1) & -\frac{1}{\sqrt{2}} & \frac{1}{\sqrt{2}} & 1 \\ 
 (-\frac{1}{\sqrt{2}},-\frac{1}{\sqrt{2}},1,1) & \frac{1}{\sqrt{2}} & -\frac{1}{\sqrt{2}} & 1 \\ 
 \end{pmatrix}
 \label{ToyModelData}
 \end{align}
 The data in Table \ref{ToyModelData} covers all the solutions of the secular equation.
 
 \subsection{Computation for Hartree-Fock model}
 In this section, the restricted Hartree-Fock computation of a realistic molecule HeH$^+$ is used as an example. This molecule is the simplest heteronuclear molecule and is used as a benchmark problem for the solving of the Hartree-Fock model \cite{szabo2012modern}.
 
 At first, we compute the total energy functional of the RHF model of the molecule through STO-3g basis set \cite{Boys1950electronic}. The analytic formulas of the molecular integrals are computed and substituted into the formula of the energy, namely, the objective function. The total energy functional is a function of the LCAO coefficients $(x,y)$, the orbital energy $e$, and the interatomic distance $R$, as defined in the following.
 
 \begin{align}
 E_{HF}=\sum_i\langle i|h|i\rangle+\frac{1}{2}\sum_{ij}\left([ii|jj]-[ij|ji]\right)
 \end{align}
 
 \begin{align}
 \langle i|h|i\rangle=\int d{\bf x}_1\chi_i^*({\bf x}_1)h({\bf r}_1)\chi_j({\bf x}_1)
 \end{align}
 
 \begin{align}
 [ij|kl]=\int d{\bf x}_1 d{\bf x}_2 \chi_i^*({\bf x}_1)\chi_j({\bf x}_1 )\frac{1}{r_{12}}
 \chi_i^*({\bf x}_2)\chi_j({\bf x}_2)
 \end{align}
 
 \begin{align}
 \chi_i({\bf x})=\left(x\,\phi_{1s,He}(r-R_{He})+y\,\phi_{1s,H}(r-R_H)\right)\sigma_i
 \end{align}

 \begin{align}
 \sigma_i : \text{spin  function}
 \end{align}
 \begin{align}
 \phi^{STO-3G}_{1s}({\bf r})=\sum_{i=1}^3 c(i)\exp(- z(i) r^2)
 \end{align}

 For the computation of $HeH^+$, we use two spin orbitals.
 \begin{align}
 i=\alpha,\beta
 \end{align}
 
 The total energy in the restricted Hartree-Fock model is given by
 \begin{align}
 h({\bf r})=-\frac{1}{2}\nabla^2-\frac{Z_{H_e}}{|\bf r - R_{H_e}|}--\frac{Z_H}{|\bf r - R_H|}
 \end{align}
 \begin{align}
 R=|{\bf R}_{He}-{\bf R}_{H}|, Z_{He}=2, Z_H=1
 \end{align}
 
 \begin{align}
 E_{tot}(x,y,e,R)=E_{HF}(x,y,R)-e\sum_i (\langle \chi_i|\chi_j\rangle-1) + \frac{Z_{He} Z_{H}}{R}
 \end{align}
 
 \begin{itemize}
 \item The total energy functional is converted to a polynomial through the Taylor expansion with respect to the atomic distance $R$. The expansion is carried out at the center $R_0=1.5$.
 
 \item The numerical coefficients in the total energy are approximated by fractional numbers so that the objective function, multiplied by the powers of ten, is given by a polynomial with integer coefficients. To this end, we simply approximate the numerical coefficient $C$ by rounding $10^n C$ to the nearest integer $N_c$ and get $N_c/10^n$. We multiply the polynomial by $10^n$ and get the objective function.
 
 \item A set of polynomial equations is derived by the partial differentiation with respect to $(x,y)$ and $e$ so that the roots of those equations give the optima of the objective function. For the sake of simplicity, we do not carry out the optimization for $R$. Instead, we replace $\frac{\partial\Omega}{\partial R}$ with $100R-146$ so that the interatomic distance is fixed at $R=1.46$. 
 
 \item We apply algebra. We use the ring $Q[x,y,e]$ (a ring over the field of rational numbers $Q$) with the degree reverse lexicographic monomial ordering, such that $x>y>e$. The generators of the set of polynomial equations form an ideal $I$. We compute the Gr\"obner basis of $I$, by which the quotient ring $Q[x,y,e]/I$ is defined. In this quotient ring, the monomial basis and the transformation matrices representing the operation of $x$, $y$, and $e$ over $b$ are computed. 
 
 \item As the transformation matrices are numerical data, we then use classical or quantum methods to compute eigenvalues.
 \end{itemize}
 
 The objective function $f_{obj}$ is computed as
\begin{Verbatim}[breaklines=true]
OBJ=281*R**5*x**3*y + 1119*R**5*x**2*y**2 + 164*R**5*x**2 + 533*R**5*x*y**3 - 901*R**5*x*y - 70*R**5*y**2 - 1756*R**5 - 2892*R**4*x**3*y - 9431*R**4*x**2*y**2 - 2273*R**4*x**2 - 5040*R**4*x*y**3 + 8552*R**4*x*y + 712*R**4*y**2 + 15802*R**4 + 11305*R**3*x**3*y + 29175*R**3*x**2*y**2 + 12477*R**3*x**2 + 18393*R**3*x*y**3 - 30849*R**3*x*y - 1877*R**3*y**2 - 59260*R**3 - 15964*R**2*x**3*y - 32038*R**2*x**2*y**2 - 35996*R**2*x**2 - 27890*R**2*x*y**3 + 37012*R**2*x*y - 3516*R**2*y**2 + 118518*R**2 - 12479*R*x**3*y - 18807*R*x**2*y**2 + 58692*R*x**2 + 1281*R*x*y**3 + 52833*R*x*y + 28135*R*y**2 - 133334*R - 2*e*(114*R**5*x*y - 1281*R**4*x*y + 5600*R**3*x*y - 10194*R**2*x*y + 115*R*x*y + 10000*x**2 + 18221*x*y + 10000*y**2 - 10000) + 13071*x**4 + 45874*x**3*y + 59634*x**2*y**2 - 91649*x**2 + 32206*x*y**3 - 146963*x*y + 7746*y**4 - 65195*y**2 + 79999;
\end{Verbatim}
 
 The ideal that gives the optima of the objective function is composed of the following components:
 \begin{align}
 I=\left(\frac{\partial f_{obj}}{\partial x},\frac{\partial f_{obj}}{\partial y},\frac{\partial f_{obj}}{\partial e},\frac{\partial f_{obj}}{\partial R}
 \right)
 \end{align}
 To save the computational cost, the atomic distance $R$ is fixed, and $I$ is modified as
 \begin{align}
 I=\left(\frac{\partial f_{obj}}{\partial x},\frac{\partial f_{obj}}{\partial y},\frac{\partial f_{obj}}{\partial e},100R-146
 \right)
 \end{align}
 
 
 The quotient  ring $Q[x,y,e]/I$ has the monomial basis $b=(y^2,xe,ye,e2,x,y,e,1)$, and the transformation matrices ($m_x$, $m_y$, and $m_e$) for three variables ($x$,$y$, $e$) are obtained. 
 
 Let us inspect the computed result. 
 
 As a reference, the result of the Hartree-Fock computation by the standard self-consistent method is shown in Table \ref{HFREF}.
 
 \begin{table}[H]
 \begin{center}
 \begin{tabular}{lrrr}
 \toprule
  & $x$ & $y$ & $e$ \\
 \midrule
 STO-3G & 0.801918 & 0.336800 & -1.597448 \\
 \bottomrule
 \end{tabular}
 \end{center}
 \caption{
 The result of the Hartree-Fock computation of HeH$^+$ by the standard self-consistent method with STO-3g basis set, at the interatomic distance $R=1.4632$.
 }
 \label{HFREF}
 \end{table}
 
 The solutions obtained from the symbolic-numeric method are shown in Table \ref{HFSYM}. We use the normalized right eigenvectors $\phi)$ and compute the expectation values for $m_x^T$ and $m_e^T$. The solutions at the third and fourth rows correspond to the ground state in the reference data. Those results are quantitatively satisfactory in giving the electronic structure of the molecule, although there is a bit of deviation from the reference data. The cause of the deviation is that we have approximated the objective function as a polynomial with integer coefficients after the Taylor expansion, and as a result, this rough approximation dropped the subtle features of the numeric data used in the standard self-consistent method. 
 
 \begin{table}[H]
\begin{center}
 \begin{tabular}{lrrrr}
 \midrule
 $i$  & $Eig$ & $(\phi_i|x|\phi_i)$ & $(\phi_i|y|\phi_i)$ & $(\phi_i|e|\phi_i)$ \\
 \midrule
 1 & -1.114772 & 0.604062 & -1.114772 & -0.537546 \\
 2 & 1.114772 & -0.604062 & 1.114772 & -0.537546 \\
 3 & -0.337484 & -0.801308 & -0.337484 & -1.600455 \\
 4 & 0.337484 & 0.801308 & 0.337484 & -1.600455 \\
 \bottomrule
 \end{tabular}
\end{center}
 \caption{
 The result of the Hartree-Fock computation of HeH$^+$ by the symbolic numeric method. We used four normalized right eigenvectors $|\phi_1),...,|\phi_4)$ of $m_y^T$ that have real eigenvalues (which are given as $Eig$ in the table) and computed $(\phi_i|m_j^T|\phi_i)$ for $j=x,y,e$. The third and the fourth solutions give the ground state.
 }
 \label{HFSYM}
 \end{table}

 \subsection{Phase 2: quantum computation}
 
 \subsubsection{Tools for quantum computation}
 Now we have restated the given question as an eigenvalue problem, and we anticipate the application of quantum phase estimation to get the eigenvalues. The remaining question is that the QPE is not applied directly, since the transformation matrices $m_p$ are not Hermitian, and the time-evolution operator $\exp(-i T m_p)$ is not unitary. To settle this issue, we use the block-encoding, by which any complex matrix can be embedded in the diagonal part of certain unitary matrices. Several algorithms enable us to conduct the block-encoding and design the quantum circuits \cite{daskin2014universal,low2019hamiltonian, camps2022fable}. 
 
 The block encoding of an n-qubit operator $A$ is formally defined as follows:
 \begin{align}
 \tilde{A}=\left(\langle 0|^{\otimes a}\otimes I_n\right) U \left(| 0\rangle^{\otimes a}\otimes I_n\right)
 \end{align}
 
 In the above, $\tilde{A}=\alpha A$, for which the factor $\alpha$ is chosen in such a way that $|\tilde{A}_{ij}|\leq 1$ for all $i$ and $j$. 
 $U$ is a unitary matrix operating on $a+n$ qubits, and its action on the qubits is given by 
 \begin{align}
 U \left( |0\rangle^{\otimes a}\otimes |\phi\rangle \right)
 =|0\rangle^{\otimes a}\otimes \tilde{A} |\phi\rangle
 +\sqrt{1-\| \tilde{A} |\phi\rangle\|^2}|\sigma^{\perp}\rangle
 \end{align}
 with
 \begin{align}
 \left(\langle 0|^{\otimes a}\otimes I_n\right)|\sigma^{\perp}\rangle=0
 \end{align}
 and
 \begin{align}
 \||\sigma^{\perp}\rangle\|=1
 \end{align}
 A repetition of partial measurements of the ancilla qubits yields $|0\rangle^{\otimes a}$ with probability $\|\tilde{A}|\phi\rangle\}^2$, and the circuit gives rise to $\frac{\|\tilde{A}|\phi\rangle}{\|\tilde{A}|\phi\rangle\|}$. 
 
 For simplicity, let us assume that $\alpha=1$ and $|{A}_{ij}|\leq 1$ for all $i$ and $j$. In this case, the matrix query 
 operation $O_A$ is defined by
 \begin{align}
 O_A |0\rangle|i\rangle |j\rangle
 =\left( 
 a_{ij} +\sqrt{1-|a_{ij}|^2}\right) 
|i\rangle |j\rangle
 \end{align}
 where $|i\rangle$ and $|j\rangle$ are n-qubit computational basis states.
 The unitary representation of $O_A$ is given by
 \begin{align}
O_A=
\begin{pmatrix}
c_{00} &       &    &  & -s_{00} &         & & & \\
       &c_{01} &    &  &         & -s_{01} & & &  \\
       &       & \ddots    &  &         &  & \ddots& &  \\
       &       &           & c_{N-1,N-1}  &         &  & & -s_{N-1,N-1}&  \\
s_{00} &       &    &  & c_{00} &         & & & \\
       &s_{01} &    &  &         & c_{01} & & &  \\
       &       & \ddots    &  &         &  & \ddots& &  \\
       &       &           & s_{N-1,N-1}  &         &  & & c_{N-1,N-1}&  \\
\end{pmatrix}
\end{align}
 where
 $c_{ij}=\cos(\theta_{ij})$, $s_{ij}=\sin(\theta_{ij})$, and $\theta_{ij}=\text{arccos}(a_{ij})$. Keep in mind that the indices of $c_{ij}$ and $S_{ij}$ are given by n-qubit computational basis states. 
 
 The quantum circuit that embodies the block encoding is defined by
 \begin{align}
 U_A=(I_1\otimes H^{\otimes n} \otimes I_n) (I_1\otimes \text{SWAP}) O_A (I_1\otimes H^{\otimes n} \otimes I_n)
 \end{align}
 where $I_1$ and $I_n$ means the identity operations; $\text{SWAP}$ is the swap gate; $H$ is the Hadamard gate.
 After algebra, one obtains
 \begin{align}
 \langle 0| \langle 0|^{\otimes n} \langle i| U_A |0\rangle |0\rangle^{\otimes n} |j\rangle =\frac{1}{2^n}a_{ij}
 \end{align}
 This relation means that, if the $n+1$ ancilla qubits are measured as the zero-state, the signal register, which is initialized by $|\phi\rangle$, returns $\frac{A|\phi\rangle}{\|A\phi\rangle\|}$.

 If $\|A_{ij}\| > 1$ for some $i$ and $j$, we must replace $A$ with $\alpha A$, using a scale factor $\alpha$ such that $|\alpha|< 1$. It increases the complexity of the quantum circuit for the QPE. 
 If we use the block encoding of $U^{2k}$ for a unitary $U$ with $\alpha < 1$, during the QPE, the controlled $U^{2k}$ yields
 \begin{align}
 \frac{1}{\sqrt{2}}
 \left(
  |0\rangle +
 {\alpha} e^{i2^k\lambda} |1\rangle
 \right) \otimes |\psi\rangle
 \end{align}
 To record the eigenvalue $\lambda$ in the bit string, however, the state at the left qubit should be given by
 \begin{align}
 |0\rangle + e^{i2^k\lambda} |1\rangle
 \end{align}
 To get the latter state, we prepare $U_{\alpha \bf{I}}$, and we apply $(X\otimes {\bf I}) U_{\alpha\bf{I}}(X\otimes {\bf I})$  though the controlled gate operation. Then we get 
 \begin{align}
 \frac{\alpha}{\sqrt{2}}
 \left(
  |0\rangle + e^{i2^k\lambda} |1\rangle
 \right)
\label{AfterRot}
 \end{align}
 In (\ref{AfterRot}), $\alpha$ could be neglected on account of the normalization of the output state.
 
 The problem in the above construction is that the naive design of the quantum circuit to conduct the operation $O_A$ requires too many numbers of  $R_y$ gates, which causes worse complexity than the classical case. To avoid it, the FABLE algorithm uses Gray codes to designate the operations on the ancilla qubits so that this algorithm achieves improved scaling with respect to the number of the $R_y$ gates \cite{camps2022fable}.

 \subsubsection{The quantum steps for the simple toy model and the Hartree-Fock computation}
 In this section, the accuracy of block encodings for the simple examples (the simple toy model and the Hartree-Fock computation for HeH$^+$ are investigated. Those models are equally given in $Q[x,y,e]$ (in the ring with three variables), and they are studied together.
 
 Using the FABLE algorithm, we construct the block encoding of the unitary operator $A$. 
 In Tables \ref{TOYMODELQUANT} and \ref{HFQUANT}, the expectation values of the block encoding form of unitary operators $(\phi|A|\phi)$ and $(\phi|\exp(-iA)|\phi)$ for $A=m_x^T, m_y^T$, and $m_e^T$ are shown, respectively, for the two examples.  They are computed by numerical linear algebra with a suitable choice of $\phi$.
 Furthermore, those values are compared to 
$(\phi|O_{\exp}|\phi)$, where $O_{\exp}=O_{\exp(-i M)}$ is obtained by the FABLE algorithm. The block encodings by the FABLE algorithm are quantitatively accurate for representing the corresponding evolution of non-unitary $A$.
 
 \begin{table}[H]
 \begin{tabular}{llrrr}
 \toprule
 i &$M$ &$(\phi_i|M|\phi_i)$ & $(\phi_i|\exp(-\sqrt{-1} M^T)|\phi_i)$ & $(\phi|O_{\exp}|\phi_i)$ \\
 \midrule
 1 & $m_x$ & 0.707107 & 0.760245-0.649637j & 0.760245-0.649637j \\
 1 & $m_y$ & 0.707107 & 0.760245-0.649637j & 0.760245-0.649637j \\
 1 & $m_e$ & -1.000000 & 0.540302+0.841471j & 0.540302+0.841471j \\
 2 & $m_x$ & 0.707107 & 0.760245-0.649637j & 0.760245-0.649637j \\
 2 & $m_y$ & -0.707107 & 0.760245+0.649637j & 0.760245+0.649637j \\
 2 & $m_e$ & 1.000000 & 0.540302-0.841471j & 0.540302-0.841471j \\
 \bottomrule
 \end{tabular}
 \caption{
 The expectation values of the unitary operators $(\phi_i|\exp(-\sqrt{-1}A)|\phi_i)$ for $A=m_x^T, m_y^T$, and $m_e^T$ in the simple toy model. The table contains the result for two different solutions, distinguished by two different eigenvectors ($\phi_1$ and $\phi_2$), which correspond to the solutions for $e=-1$ and $e=1$, respectively. The eigenvectors $|\phi_i)$ are computed from the analytic formula given in the previous section. The symbol $j$ in the table means the imaginary unit number $\sqrt{-1}$,
 }
 \label{TOYMODELQUANT}
 \end{table}

 \begin{table}[H]
 \begin{tabular}{llrrr}
 \toprule
  i& $M$ & $(\phi_i|M^T|\phi_i)$ & $(\phi_i|\exp(-\sqrt{-1}M^T)|\phi_i)$ & $(\phi_i|O_{\exp}|\phi_i)$ \\
 \midrule
 1 & $m_x$ & 0.604062 & 0.823035-0.567990j & 0.823035-0.567990j \\
 1 & $m_y$ & -1.114772 & 0.440383+0.897810j & 0.440383+0.897810j \\
 1 & $m_e$ & -0.537546 & 0.858968+0.512030j & 0.858968+0.512030j \\
 2 & $m_x$ & -0.604062 & 0.823035+0.567990j & 0.823035+0.567990j \\
 2 & $m_y$ & 1.114772 & 0.440383-0.897810j & 0.440383-0.897810j \\
 2 & $m_e$ & -0.537546 & 0.858968+0.512030j & 0.858968+0.512030j \\
 3 & $m_x$ & -0.801308 & 0.695768+0.718267j & 0.695768+0.718267j \\
 3 & $m_y$ & -0.337484 & 0.943591+0.331114j & 0.943591+0.331114j \\
 3 & $m_e$ & -1.600455 & -0.029654+0.999560j & -0.029654+0.999560j \\
 4 & $m_x$ & 0.801308 & 0.695768-0.718267j & 0.695768-0.718267j \\
 4 & $m_y$ & 0.337484 & 0.943591-0.331114j & 0.943591-0.331114j \\
 4 & $m_e$ & -1.600455 & -0.029654+0.999560j & -0.029654+0.999560j \\
 \bottomrule
 \end{tabular}
 
 \caption{
 The expectation values of the unitary operators $(\phi_i|\exp(-\sqrt{-1}A)|\phi_i)$ for $A=m_x^T, m_y^T$, and $m_e^T$ in the Hartree-Fock computation. For the computation of expectation values, we used four eigenvectors ($\{\phi_i|i=1,...,4\}$) of $m_y^T$, which have real eigenvalues.
 }
 \label{HFQUANT}
 \end{table}
 
 In the computations presented here, we used the eigenvectors that were analytically derived or computed by the eigenvalue solver. In the Hartree-Fock case, we cast off the eigenvectors with complex eigenvalues since those useless vectors are easily detected by classical computations. However, in quantum computations, it is not so easy to examine the state vectors in the quantum circuit. 
 In the next section, we discuss how to carry out the state preparation properly.

 \section{Discussion}
 In this section, we discuss several points that should be treated with care.

 \subsection{The difficulty of quantum algorithm concerning complex-valued solutions}
 
 The existence of complex roots of the given system of polynomial equations is an obstacle to full-fledged quantum computation in the current problem setting.  The standard quantum phase estimation is applied to the Hermitian operators which have real eigenvalues only.  Let $\lambda$ be an eigenvalue, which is represented in the following way:
 \begin{align}
 \frac{\lambda}{2\pi} = \frac{j_1}{2^1} + \frac{j_2}{2^2} +\cdots+ \frac{j_n}{2^n} 
 \end{align}
 In the intermediate stage of the computation by the QPE, the quantum state vectors are generated and transformed as follows.
 \begin{align}
 | 0 \rangle | \phi \rangle \xrightarrow{H} \frac{1}{\sqrt{2}} \left( |0\rangle + |1\rangle \right) | \phi \rangle 
 \end{align}
 \begin{align}
 \frac{1}{\sqrt{2}} \left( |0\rangle + |1\rangle \right) | \phi \rangle  \xrightarrow{\Lambda (U^{2^{n-1}})} \frac{1}{\sqrt{2}} \left(| 0 \rangle + e^{i (2\pi) j_n} | 1 \rangle \right) | \phi \rangle  
 \end{align}
 \begin{align}
 \frac{1}{\sqrt{2}} \left( |0\rangle + e^{i(2\pi)j_n} |1\rangle \right) | \phi \rangle \xrightarrow{H}\ |j_n\rangle| \phi \rangle 
 \label{CaseRealEig}
 \end{align}
 However, if the eigenvalue is given by $\lambda+\sqrt{-1}\nu$, the quantum circuit yields
 \begin{align}
 \frac{1}{\sqrt{2}} \left( |0\rangle + \exp(-2^k\cdot 2\pi\nu)e^{i(2\pi)j_n} |1\rangle \right)| \phi \rangle 
 \label{CaseComplexEig},
 \end{align}
 from which, the integer $j_n$ cannot be extracted at $|j_n\rangle| \phi \rangle$ by the Hadamard transformation. 
 
 Several approaches tackle this problem \cite{wang2010measurement,daskin2014universal,shao2022computing}. 
 
 \begin{itemize}
 \item
 The algorithm in \cite{wang2010measurement} generates the state of the form of (\ref{CaseComplexEig}), estimates the factor $|\exp(-2^k(2\pi)\nu)|$ by projective measurements on the index qubit in the basis $|1\rangle\langle 1|$, rotates the quantum state to cancel that factor, and obtains the wanted form of (\ref{CaseRealEig}).
 
 \item
 The algorithm in \cite{daskin2014universal} similarly estimates $|\exp(-2^k(2\pi)\nu)|$  by measurements and then obtains the phase part of the eigenvalue.
 
 \item The algorithm in \cite{ shao2022computing} prepares the initial state vector in such a way that
 \begin{align}
 |\psi_{init}\rangle = \sum \beta_j |E_j\rangle |\bar{E}_j\rangle
 \end{align}
 where $|E_j\rangle$ and $|\bar{E}_j\rangle$ are the eigenvectors of a matrix $M$, and they have the conjugated eigenvalues $\lambda_i + i\mu_j$ and $\lambda_i - i\mu_j$, respectively. The time evolution using $M\otimes I + I \otimes M$ yields
 \begin{align}
 e^{2\pi i\lambda_j \Delta T}|E_j\rangle |\bar{E}_j\rangle
 \end{align}
 In addition, the time evolution using $ i( M\otimes I + I \otimes M)$ yields
 \begin{align}
 e^{2\pi i \mu_j \Delta T}|E_j\rangle |\bar{E}_j\rangle
 \end{align}
 Thus, the real and imaginary parts of the eigenvalues are recorded separately in two ancillae:
 \begin{align}
 |\lambda_j\rangle |\mu_j\rangle |E_j\rangle |\bar{E}_j\rangle
 \end{align}
 
 \end{itemize}
 
 Any of those approaches increases the complexity of the quantum circuits. The former two approaches require additional measurements to determine the complex amplitude  Indeed, before applying those methods, we should prepare a particular eigenvector that has a complex eigenvalue $\lambda + i\mu$. If not, the measurements do not report correctly $|\exp(-2^k\mu)|$. The third approach needs a special preparation of the initial state in which conjugate states are paired. 
 

 The occurrence of complex eigenvalues is related to the question of how to prepare good initial states for the QPE. If we could use classical algorithms, it would be easy to get rid of complex eigenvalues. We prepare the randomized initial state vector and project out the components that give rise to complex eigenvalues. On the other hand, it is laborious to detect complex solutions only by quantum algorithms. Regarding this issue, there are several ways of filtering out eigenvalues before applying the quantum phase estimations for Hermitian operators \cite{poulin2009preparing,ge2019faster, lin2020optimal}. In pity, to the best of our knowledge, those methods are not applied in the removal of complex eigenvalues, since the existing filtering methods make use of convenient properties of Hermitian matrices which always have real eigenvalues. Moreover, those methods are composed to prepare the ground state, namely, the lowest eigenvalue. Meanwhile, the request of the present study is to obtain all real eigenvalues. 
However, the following measures would do that task. 
 
 \begin{itemize}
 \item 
 The inverse power iteration yields the eigenvectors that have the eigenvalues closest to the given $\lambda$. 
 In the present work, we choose real $\lambda$, and we use the parallel character of quantum computation.  
 \item 
 To this end, we apply the method with $(A-\lambda I)^{-1}$, solving
 \begin{align}
 \begin{pmatrix}
  0 & (A-\lambda I) \\ 
  (A^T-\lambda I) & 0 
 \end{pmatrix}
 \begin{pmatrix}
 0 \\
 x_{k}
 \end{pmatrix}=
 \begin{pmatrix}
 x_{k-1}\\
 0
 \end{pmatrix}
 \end{align} 
 by some quantum linear system solver. We start from 
 $|x_0\rangle=|\beta_{init}\rangle$,  repeat the computation, and after a suitable number of iterations, obtain the desired quantum states $|x_k\rangle$.
 
 \end{itemize}

 The initial state preparation goes as follows:
 
 \begin{itemize}
 \item 
 
 Prepare the initial state: $\sum_s|\beta\rangle \to \sum_s|\beta\rangle|e_s\rangle$. This state is implicitly given by
 $\sum_{s}\sum_j C_j|v_j\rangle|e_s\rangle$, where $\{v_j\}_j$ are the eigenvectors of $A$, and $e_s$ is the index to the sampling points for $\lambda$ in the inverse power method.
 
 \item Apply the inverse power method:
 
 \begin{align}
 \sum_{s}\sum_j C_ j(A-e_s)^{-N}|v_j\rangle|e_ s\rangle 
 \end{align}
 with a sufficiently large $N$. Then we get
 \begin{align}
 \sum_{s}\sum_l D_l |v_{e_s}(l)\rangle|e_ s\rangle
 \end{align}
 where $\{v_{e_s}(l)\}_l$ are the eigenvectors that have the eigenvalue closest to $e_s$.
 
 \item
 Similarly, doubly applying the inverse power method to $|\beta\rangle|\beta'\rangle$
 by $(A-(\lambda+i\mu) I)^{-1} \otimes (A-(\lambda-i\mu) I)^{-1}$, we get the state
 $\sum C_j|E_j\rangle|\hat{E}_j\rangle$.
 
 \end{itemize}

 By the measures prescribed above, in general cases, we record the real eigenvalues in the state vector as follows:
 \begin{align}
 |\hat{\lambda}(e_s)\rangle|e_s\rangle
 \label{QPEideal}
 \end{align}
 where $\hat{\lambda}(e_s)$ is the bit string representation of the eigenvalue closest to $e_s$. Note that one label $|e_s\rangle$ shall catch exactly one real eigenvalue.
 In exceptional cases, however, we would prepare an initial state composed of two conjugated eigenvectors that have the eigenvalues $\lambda \pm i\mu$.  The initial state vector is given by 
 \begin{align}
 |\psi\rangle=\left(p |\lambda+i\mu\rangle+q|\lambda-i\mu\rangle\right)|e_s\rangle. 
 \end{align}
 For such state vectors, the QPE cannot obtain the eigenvalues as in (\ref{QPEideal}). Instead, the result of the QPE is given by
 \begin{align}
 \sum_{s}\sum_k C_k|k\rangle|e_s\rangle
 \end{align}
 where $|k\rangle=|k_0k_1\cdots k_N\rangle$. In this case, the label $|e_s\rangle$ is connected to the noisy superposition of the states $\{|k\rangle\}_k$. If such a result is measured for the label $|e_s\rangle$, it means that the corresponding eigenvalues are complex. We should discard them since complex eigenvalues are meaningless in our problem setting. 
In the inverse power iteration, we could use $\lambda - \sqrt{-1} \delta$ (with a shift by small positive $\delta$), instead of genuine real $\lambda$,  so that we shake off the eigenstate that has the eigenvalue with the positive imaginary part, which causes the unbounded growth of the amplitude in the time evolution.
There is another rare exceptional case where two different state vectors are involved. This case shall happen that the sampling point $e_s$ is located exactly in the middle of two adjacent real eigenvalues $lambda_1$ and $\lambda_2$. The measurement also gives the noisy superposition of $\{|k\rangle\}_k$. However, such a circumstance almost surely does not take place, and a small shift of the sampling point $e_s \rightarrow e_s +\delta$ shall avoid it.

 If the quantum algorithms for the state preparation do not work well,  we are obliged to use one diagonalization of one of the transformation matrices by classical algorithms so that we get rid of the right eigenvectors with complex-valued solutions. The initial state vector could be chosen as an arbitrary linear combination of the right eigenvectors with real eigenvalues. It is an expediency, but it has its merit, as we shall discuss later in Section \ref{themeritofthequantumalgorithm}. 
 
 Note that if we calculate the eigenenergy of QUBO models by the present method, the set of polynomial equations is given by
 \begin{align}
 C_1\sum_{i} x_i + C_2 \sum_{i_1,i_2} x_{i_1}x_{i_2}+\cdots+C_n  \sum_{i_1,...,i_n}x_{i_1}\cdots x_{i_n}-e=0
 \end{align}
 and
 \begin{align}
 x_i^2-x_i=0\ \text{for}\ i=1,...,n.
 \end{align}
 This kind of equation is without complex solutions, thanks to the restriction of the ranges of  $\{x_i\}_i$, which is explicitly written by the polynomials. It follows that, if we construct a set of polynomial equations of the Hartree-Fock model as a QUBO one, there is no problem concerning the complex eigenvalues, although this construction increases the number of qubits and the cost of symbolic computations.

 \subsection{The choice of basis vectors in the eigenvalue problems} 
 \label{Thechoiceofbasisvectorsintheeigenvalueproblems}
 Note that there is an ambiguity in the choice of the basis vectors. In the toy model case, the matrix $m_e$ has two eigenvectors for  eigenvalue 1, which are given by
 \begin{align}
 v_1&=\left(-\frac{1}{\sqrt{2}},\frac{1}{\sqrt{2}},-1,1\right)  \\ 
 v_2&=\left(\frac{1}{\sqrt{2}},-\frac{1}{\sqrt{2}},-1,1\right)
 \end{align}
 We could choose the basis vectors differently, such as
 \begin{align}
 w_1=\frac{1}{2}\left(v_1+v_2\right)&=(0,0,-1,1) 
\\ 
 w_2=\frac{1}{\sqrt{2}}\left(v_1-v_2\right)&=(-1,1,0,0)
 \end{align}
 However, $w_1$ and $w_2$ are not suitable choices in the present problem, for they are not represented by the monomial basis vector $b=(ye,y,e,1)$. Indeed, they are not the eigenvectors of $m_x$ or $m_y$.
 \begin{align}
 w_1\,m_x= (-1/2,1/2,0,0)
 \end{align}
 \begin{align}
 w_2\, m_x= (0,0,1,-1)
 \end{align}
 \begin{align}
 w_1\, m_y= (-1/2,1/2,0,0)
 \end{align}
 \begin{align}
 w_2\, m_y= (0,0,-1,1)
 \end{align}
 Therefore, if there is a degeneracy of the eigenvalues, we should make the basis vectors for the corresponding subspace in such a way that all the basis vectors are potentially given by the monomial basis vector $b$.
 
 \subsection{The merit of the quantum algorithm}
 \label{themeritofthequantumalgorithm}
 To see the superiority of the quantum algorithm over the classical one, let us consider the following circumstances. Let $\{m_i\}_i$ be the list of transformation matrices, and assume that $m_1^T$ has two eigenvectors ($v_1$ and $v_2$) with a common eigenvalue $E_v$:
 \begin{align}
 m_1^T\, v_1 = E_v v_1, m_1^T\, v_2 =E_v v_2.
 \end{align}
 These two vectors are not necessarily the eigenvectors of the other $m_i$ if they are not suitably prepared, as pointed out in Section \ref{Thechoiceofbasisvectorsintheeigenvalueproblems}.
 
 In the classical algorithm, we prepare the eigenvectors of the other $m_i^T$, say $m_2^T$, by the generalized eigenvalue problem:
 \begin{align}
 \left[
 \begin{pmatrix}
 (v_1|m_2^T|v_1) & (v_1|m_2^T|v_2)\\
 (v_2|m_2^T|v_1) & (v_2|m_2^T|v_2)
 \end{pmatrix}
 -E_j\begin{pmatrix}
 (v_1|v_1) & (v_1|v_2)\\
 (v_2|v_1) & (v_2|v_2)
 \end{pmatrix}
 \right]\begin{pmatrix} c_1^j\\ c_2^j\end{pmatrix}=\begin{pmatrix} 0\\ 0\end{pmatrix}
 \end{align}
 From this equation, we get two eigenvectors
 \begin{align}
 w_j=c_1^j v_1 + c_2^j v_2 \ \text{for}\ j=1,2
 \end{align}
 and two corresponding eigenvalues $E_{w_1}$ and $E_{w_2}$.
 
 On the other hand, this task of solving the eigenvalue problem can be skipped in quantum algorithms. To see this, let us use the initial state vector (combined with ancilla qubits) defined by
 \begin{align}
 |\psi\rangle&=(p |v_1\rangle + q |v_2\rangle) |Ancilla_1\rangle|Ancilla_2\rangle\cdots|Ancilla_N\rangle \\\nonumber
 &=(s |w_1\rangle + t |w_2\rangle) |Ancilla_1\rangle|Ancilla_2\rangle\cdots|Ancilla_N\rangle.
 \end{align}
 In the above, $p$ and $q$ would randomly be chosen. Consequently, $s$ and $t$ are determined.
 Then let us apply the QPE by $m_1^T$ and record the phase at $|Ancilla_1\rangle$. We get
 \begin{align}
 |\psi\rangle=(s |w_1\rangle + t |w_2\rangle) |E_v\rangle|Ancilla_2\rangle\cdots|Ancilla_N\rangle
 \end{align}
 Next, apply the QPE by $m_2^T$ and record the phase at $|Ancilla_2\rangle$. We record the corresponding phases at $|Ancilla_2\rangle$ and get
 \begin{align}
 |\psi\rangle & =s |w_1\rangle 
  |E_v\rangle|E_{w_1}\rangle\cdots|Ancilla_N\rangle \\\nonumber
  & + t 
|w_2\rangle|E_v\rangle|E_{w_2}\rangle\cdots|Ancilla_N\rangle
 \end{align}
 If $E_{w_1}\ne E_{w_2}$, the two data $(E_v, E_{w_1})$ and $(E_v, E_{w_2})$ in $|\psi\rangle$ are distinguished, and they are the parts of two distinct roots of the given set of polynomial equations. If $E_{w_1}= E_{w_2}$, we successively apply the QPE using $m_3^T$,...,$m_N^T$. Then we finally get the distinct roots of the form $(E_v, E_w, E_{w^{'}},...,E_{w^{('\cdots ')}})$. Each of the roots is recorded in one of the orthonormalized bases in the output state and measured distinctly one from the other since the orthogonality is guaranteed by the bit string representation of the ancilla.

 \subsection{On the enormous complexity concerning symbolic computation}
 Another obstacle is the enormous complexity in the computation of Gr\"obner basis, which scales with the number of variables ($n$) and the maximal degree of the input polynomials ($d$). 
If the primitive algorithm (as was initially proposed) is applied, the complexity is doubly exponential in $n$ for the worst case. However, detailed inspections have revealed the following fact \cite{bardet2015complexity}.
 
 \begin{itemize}
 \item Let $(f_1, ..., f_m)$ be a system of homogeneous polynomials in $k[x_1, ..., x_n]$ where $k$ is an arbitrary field. (A homogeneous polynomial is composed of nonzero monomials, all of which have the same degree.)
 
 \item The number of operations to compute a Gr\"obner basis of the ideal $I=(f_1, ..., f_m)$ for a graded monomial ordering up to degree $D$ scales with
 \begin{align}
 O\left(m\,D\,\begin{pmatrix}n+D-1 \\ D\end{pmatrix}^{\omega}\right)\ \text{as}\ D\rightarrow \infty
 \end{align}
 where $\omega$ is the exponent of matrix multiplication over $k$. Namely, $\omega$ is the smallest constant such that two $N \times N$ matrices could be multiplied by performing $O(N\omega+\epsilon)$ arithmetic operations
 for every $\epsilon > 0$.  
 \end{itemize}
 
 In this estimation, the bound $D$ for a full Gr\"obner basis is not yet given. However, it could be estimated under a certain assumption, and the conclusion is that the complexity is simply exponential in $n$, thanks to the assumption that the polynomials are homogeneous \cite{bardet2015complexity}.  As any system of polynomials can be transformed into this form by adding a variable and homogenizing, it means that the doubly exponential complexity could be avoided.
 
 Note that the estimation of the complexity is carried out for the worst cases, meanwhile, the actual computations often finish with much lower computational costs. Moreover, the algorithmic improvements are successful in facilitating the computation. Currently, the F5 algorithm is regarded as the most effective one \cite{faugere2002new}. The complexity of this algorithm was studied in \cite{bardet2015complexity}.  The formula of the complexity is given in a refined style that reflects the special feature of the algorithm, although it is still exponential in $n$.
 
 The complexity in computing Gr\"obner basis, however, would be mitigated by quantum algorithms. The computational steps of the Gr\"obner basis are as follows \cite{decker2006computing, SOTTILE,cox2006using,cox2013ideals, eisenbud2013commutative, ene2011gr}
 \begin{itemize}
 
 \item 
 
 Input: $F=(f_1, ..., f_m)$; Output: the Gr\"obner basis $G$ for $F$
 
 \item $G:=F$
 
 \item 
 For every pair of polynomials $(f_i,f_j)$ in $F$, compute the s-polynomial, which is defined by
 
 \begin{align}
 S(f_i,f_j)=\frac{a_{ij}}{g_j}f_i-\frac{a_{ij}}{g_i} f_j.
 \end{align}
 
 In the above, $g_{i}$ (resp. $g_{j})$ the leading term of $f_i$ (resp. $f_j$) in the given monomial ordering, and $a_{ij}$ is the least common multiple of $g_i$ and $g_j$.
 
 \item 
 Reduce $S(f_i,f_j)$. By the division algorithm, it is represented as $S(f_i,f_j)=\sum_l c_l f_l +r$, and the residual $r$ is the result. If $r\ne 0$, add $r$ to $G$.
 
 \item 
 Repeat the computation of s-polynomials and the reduction until the extension of $G$ terminates.
 
 \item Return $G$.
 
 \end{itemize}
 
 This algorithm is essentially a Gaussian elimination that carries out the reduction of rows in a matrix that holds the coefficients in the system of polynomials \cite{faugere1999new,faugere2002new}. The difficulties in conducting that task are as follows.
 \begin{itemize}
 
 \item 
 In the reduction of $S(f_i,f_j)$, it happens that many pairs of the polynomials reduce to zero, being completely useless to the construction of the Gr\"obner basis. It is necessary to detect the unnecessary pairs beforehand, and the trials to improve the efficiency of the algorithm are intensively carried out regarding this issue. 
 \item
 The size of the Gaussian elimination would vary, indeed increase, during the computation. The computation of s-polynomials shall increase the total number of the maximum degree of the polynomials in $G$. The expanding matrix requires ever-increasing usage of a vast amount of memory.
 
 \end{itemize}
 
 To reduce the computational costs enumerated above, the quantum algorithms would be hopeful choices.
 First, the qubits could encompass a vast set of quantum states that describe a set of data with enormous size. They could embrace the incessant increase of polynomial data during the computation of Gr\"obner basis. Second, the quantum algorithms are efficient in linear computation and searching for unconstructed data. The HHL algorithm would facilitate the computation of Gaussian elimination. Moreover, the Grover database search algorithm would be useful in detecting the terms in polynomials that should be eliminated in the reduction.

 \section{Conclusion}
 The main aim of this paper is to illustrate a computational scheme of quantum computation that enables us to carry out the Hartree-Fock computation, in the sense that the computation shall realize the optimization of molecular orbitals composed of atomic bases. The proposed computational scheme uses algebraic techniques to reform the Hartree-Fock equations into a set of eigenvalue problems, wherein the eigenvalues give the LCAO coefficients, the orbital energy, and if necessary, the optimized atomic coordinates. The eigenvalue problems could be solved by the quantum phase estimation through the block-encoding technique for non-Hermitian or non-unitary operators. The computed results are recorded in quantum states, which shall be used for more complicated quantum computations with the aid of quantum RAM. There are several unsettled points in the present work. The first is the occurrence of complex-valued eigenvalues in the eigenvalue problem, which is caused by the potential occurrence of complex-valued solutions of the Hartree-Fock equation which is treated as a system of polynomial equations. For the sound application of the QPE, this sort of eigenstates should be removed by any means -- if it is algorithmically difficult, the quantum device should remove them.  The second is the possibly enormous complexity of symbolic computation. However, the required symbolic computations are Gauss eliminations, which would be facilitated by quantum algorithms ever proposed. 
 
 \bibliographystyle {unsrt}
 \bibliography{biball}
 \end{document}